\documentclass[journal]{IEEEtran}

\usepackage{color}
\usepackage{lineno}
\usepackage{cite}
\usepackage{verbatim}
\usepackage{pifont}
\usepackage{stfloats}
\usepackage{amsmath,amssymb,amsfonts}
\usepackage{algorithmic}
\usepackage{graphicx}
\usepackage{textcomp}
\usepackage{xcolor}
\usepackage{subfigure}
\usepackage{setspace}
\usepackage{svg}
\usepackage[ruled]{algorithm2e}
\usepackage{booktabs}
\usepackage{url}
\usepackage{diagbox}
\usepackage{xcolor}
\usepackage{multicol}
\usepackage{multirow}
\usepackage{tikz}
\usepackage[colorlinks=true,linkcolor=blue,citecolor=blue,urlcolor=blue]{hyperref}
\definecolor{lime}{HTML}{A6CE39}
\DeclareRobustCommand{\orcidicon}{%
    \begin{tikzpicture}
    \draw[lime, fill=lime] (0,0) 
    circle [radius=0.16] 
    node[white] {{\fontfamily{qag}\selectfont \tiny ID}};    \draw[white, fill=white] (-0.0625,0.095) 
    circle [radius=0.007];    \end{tikzpicture}
    \hspace{-2mm}}
\foreach \x in {A, ..., Z}{%
    \expandafter\xdef\csname orcid\x\endcsname{\noexpand\href{https://orcid.org/\csname orcidauthor\x\endcsname}{\noexpand\orcidicon}}
    }

\SetCommentSty{mycommfont}

\SetKwInput{KwInput}{Input}                
\SetKwInput{KwOutput}{Output}              

\begin{document}



\title{Exploration of Low-Cost but Accurate Radar-Based Human Motion Direction Determination\\
\thanks{Manuscript received XXXXXXX XX, 2025; revised XXXXXXX XX, 2025; accepted XXXXXXX XX, 2025. Date of publication XXXXXXX XX, 2025; date of current version XXXXXXX XX, 2025.\par
My Bio: My name is Weicheng Gao. I'm a Ph.D. student from Beijing Institute of Technology. I’m majored and interested in mathematical and modeling theory research of signal processing, radar signal processing techniques, and AI for radar, apprenticed under professor Xiaopeng Yang. I’m currently dedicated in the field of Through-the-Wall Radar Human Activity Recognition. Looking forward to learning and collaborating with more like-minded teachers and mates. (e-mail: JoeyBG@126.com).\par
Digital Object Identifier 10.48550/arXiv.2508.XXXXX.\par}}

\author{Weicheng~Gao\orcidA{},~\IEEEmembership{Graduate~Student~Member,~IEEE}   
        \vspace{-0.6cm}
        }
        
\markboth{arXiv Preprint, August, 2025}%
{Shell \MakeLowercase{\textit{et al.}}: Bare Demo of IEEEtran.cls for IEEE Journals}

\maketitle

\begin{abstract}
This work is completed on a whim after discussions with my junior colleague. The motion direction angle affects the micro-Doppler spectrum width, thus determining the human motion direction can provide important prior information for downstream tasks such as gait recognition. However, Doppler-Time map (DTM)-based methods still have room for improvement in achieving feature augmentation and motion determination simultaneously. In response, a low-cost but accurate radar-based human motion direction determination (HMDD) method is explored in this paper. In detail, the radar-based human gait DTMs are first generated, and then the feature augmentation is achieved using feature linking model. Subsequently, the HMDD is implemented through a lightweight and fast Vision Transformer-Convolutional Neural Network hybrid model structure. The effectiveness of the proposed method is verified through open-source dataset. The open-source code of this work is released at: \href{https://github.com/JoeyBGOfficial/Low-Cost-Accurate-Radar-Based-Human-Motion-Direction-Determination}{Github/JoeyBGOfficial/Radar-Based-HMDD}.\par
\end{abstract}

\begin{IEEEkeywords}
mmWave radar, human motion determination, micro-Doppler signature, neural networks.
\end{IEEEkeywords}

\IEEEpeerreviewmaketitle

\section{Introduction}
\IEEEPARstart{R}{adar-based} human motion direction detection (HMDD) enables real-time monitoring and understanding of human gait characteristics \cite{Main1, Main2}, providing support for applications in smart homes and security surveillance \cite{Main3, Main4}. As a component of radar-based human gait recognition, it holds significant research value \cite{Main5, Main6, Main7,Main8}.\par
Researches of radar-based HMDD were widely carried out internationally. A method for quantitatively evaluating the suitability of different data domains for representing human motion was proposed by Jokanović et al. \cite{Jokanovic}. A dynamic range–Doppler-trajectory approach was put forward by Ding et al. \cite{Ding}. High-accuracy HMDD based on micro-Doppler signature combined with bio-inspired feature extraction was achieved by Song et al. \cite{Song}. A human-activity classification method that combined a multistatic radar system with motion-direction estimation was proposed by Qiao et al. \cite{Qiao}. An omnidirectional few-shot motion-recognition algorithm with good generalization performance was developed by Yang et al. \cite{Yang}. Other influential works in this field included \cite{Qi, Jeon, Chang, Guendel, Waqar}. Higher accuracy and suitability in more complex scenario configurations were mainly achieved by these methods.\par
Our team is currently working on radar-based human gait recognition. Through our findings, the effectiveness of gait feature extraction is highly affected by the human motion direction, which is an important physical factor of the micro-Doppler spectrum width. HMDD can be considered as an integral part of gait recognition. Solutions with low cost and high accuracy performance are still considered to have room for further research \cite{HMDD Drawbacks}.\par
Based on the discussions above, in this paper, a low-cost but accurate radar-based HMDD method is explored. The feature linking model (FLM) is improved to enhance micro-Doppler signature and a hybrid Vision Transformer (ViT)-Convolutional Neural Network (CNN) architecture is used to achieve direction recognition. The effectiveness of the proposed method is validated through open-source dataset.\par
\begin{figure}
    \centering
    \includegraphics[width=0.48\textwidth]{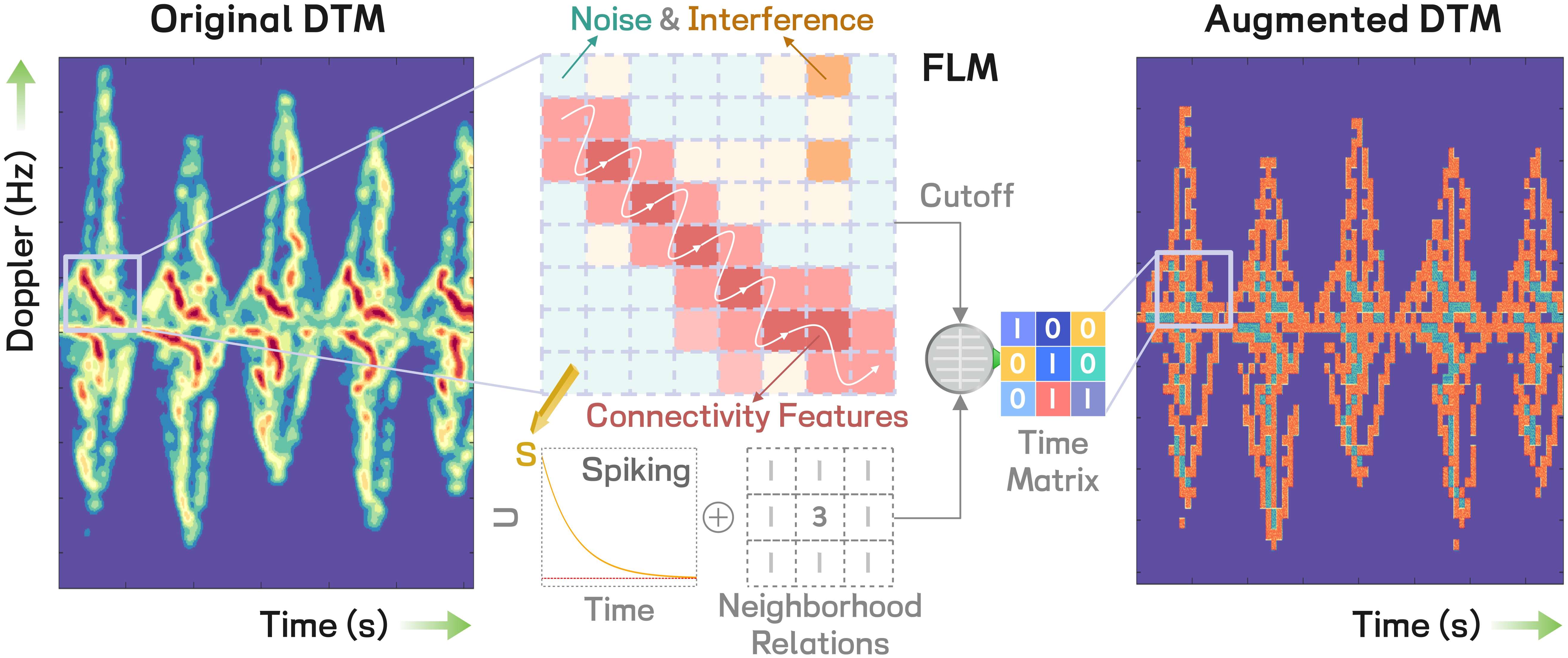}
    \caption{Schematic of the proposed FLM-based feature augmentation method.}
    \label{FLM}
    \vspace{-0.4cm}
\end{figure}\par

\section{Micro-Doppler Signature Augmentation}
Based on the open-source dataset in \cite{Dataset}, four-channel non-coherent accumulation, pulse compression, high-pass filtering, range cell selection, and short-time Fourier transform (STFT) are performed on radar human walking echoes in different motion directions to obtain Doppler-time maps (DTMs). The micro-Doppler signature on DTM is augmented using FLM.\par
\begin{figure*}
    \centering
    \includegraphics[width=\textwidth]{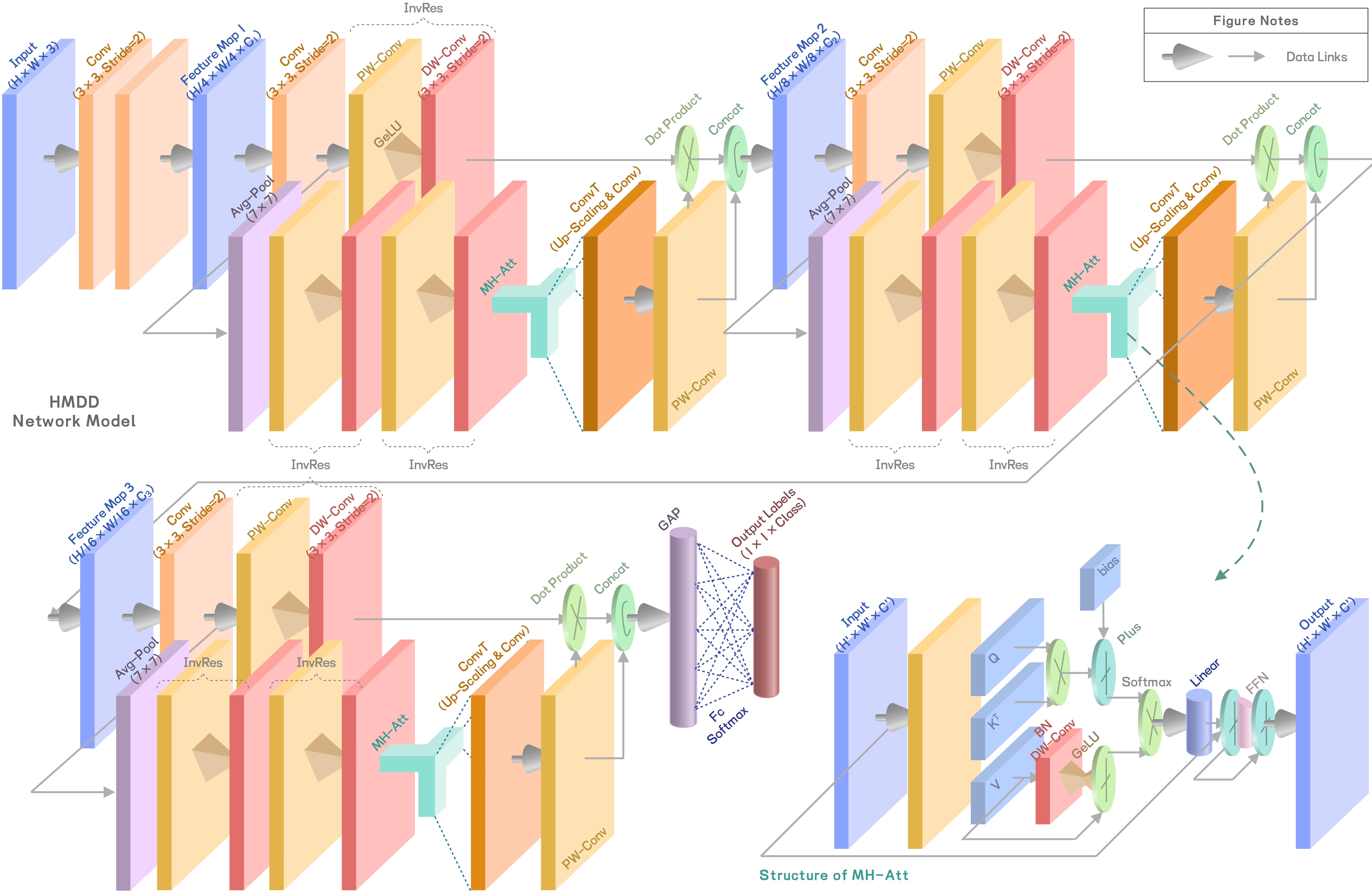}
    \caption{Schematic of the proposed low-cost accurate HMDD model based on a ViT-CNN hybrid architecture.}
    \label{HMDD_Network}
    \vspace{-0.2cm}
\end{figure*}\par
As shown in Fig. \ref{FLM}, FLM is essentially a recursive spiking neural network model \cite{FLM}. Define time matrix $T$ to represent the time step at which each neuron first discharges, $\xi$ to be the number of discharged neurons, $n=0$ to be the initial time step, $U_{ij}(n)$ to be the membrane potential and $U_{ij}(0)=0$, and $\theta_{ij}(n)$ to be the threshold, where $i,j$ are the horizontal and vertical coordinate indexes of the image. For each time step $n=n+1$ loop execution, the membrane potential is:

\vspace{-0.4cm}
\begin{equation}
U_{ij}(n) = f U_{ij}(n - 1) + S_{ij} + \alpha F_{ij}(n) + \beta L_{ij}(n),
\end{equation}
where $f\in(0,1)$ is the decay constant, $\alpha,\beta>0$ are the weights of feeding and linking input, respectively, $S_{ij}$ is the image pixel intensity. Feeding input in this work is defined as:

\vspace{-0.2cm}
\begin{equation}
F_{ij}(n) \equiv S_{ij}.
\end{equation}\par
Linking input represents the effect of neighboring neurons on the current neuron:

\vspace{-0.2cm}
\begin{equation}
L_{ij}(n) = \sum_{k,l \in N_{ij}} W_{kl,ij} Y_{kl}(n - 1) - \epsilon,
\end{equation}
where $N_{ij}$ is the neighborhood of neuron $(i,j)$, $W_{kl,ij}$ is the connection weight from neuron $(k,l)$ to $(i,j)$, expressed as the inverse of the Euclidean distance, $\epsilon$ is a small positive global suppression term. The action potential is defined as:

\vspace{-0.3cm}
\begin{equation}
Y_{kl}(n-1) = \begin{cases} 
1 & \quad \text{If} ~U_{kl}(n-1) > \theta_{kl}(n-1) \\
0 & \quad\text{Otherwise}
\end{cases}.
\end{equation}\par
Thus the recursive equation for the membrane potential is:

\vspace{-0.2cm}
\begin{equation}
\begin{aligned}
U_{ij}(n) &= f U_{ij}(n - 1) + (1 + \alpha) S_{ij} \\&+ \beta \sum_{k,l \in N_{ij}} W_{kl,ij} Y_{kl}(n - 1) - \beta \epsilon.
\end{aligned}
\end{equation}\par
Then the threshold is updated:

\vspace{-0.2cm}
\begin{equation}
\theta_{ij}(n) = g \theta_{ij}(n - 1) + h Y_{ij}(n - 1),
\end{equation}
where $g\in(0,1),h>0$ are the threshold decay constant and threshold amplification factor, respectively. If $Y_{ij}(n) = 1$ and $T_{ij} = 0$, then update $T_{ij}=n$ and $\xi = \xi+1$. The iteration is stopped when $\xi$ reaches the total number of neurons.\par
The micro-Doppler signature augmentation is achieved with the help of the final time matrix $T$ obtained above. The normalization of DTM is first achieved:

\vspace{-0.2cm}
\begin{equation}
S_{ij} = \frac{\mathrm{DTM}_{ij} - \min(\mathrm{DTM})}{\max(\mathrm{DTM}) - \min(\mathrm{DTM})} + \varepsilon,
\end{equation}
where $\max(\cdot),\min(\cdot)$ Indicates that the largest/smallest element of the matrix is found, respectively, $\varepsilon>0$ is a small bias to ensure $S_{ij}>0$. Reverse the time matrix $T$ to obtain:

\vspace{-0.2cm}
\begin{equation}
R_{ij} = \max(T) + 1 - T_{ij}.
\end{equation}\par
Finally the augmented DTM $J$ is obtained:

\vspace{-0.2cm}
\begin{equation}
J_{ij} = \left \lfloor 255\cdot \frac{R_{ij} - \min(R)}{\max(R) - \min(R)} + 0.5 \right \rfloor
\end{equation}\par
The effectiveness can be demonstrated by deriving the relationship between the time matrix and the input stimulus.\par
\begin{figure*}
    \centering
    \includegraphics[width=\textwidth]{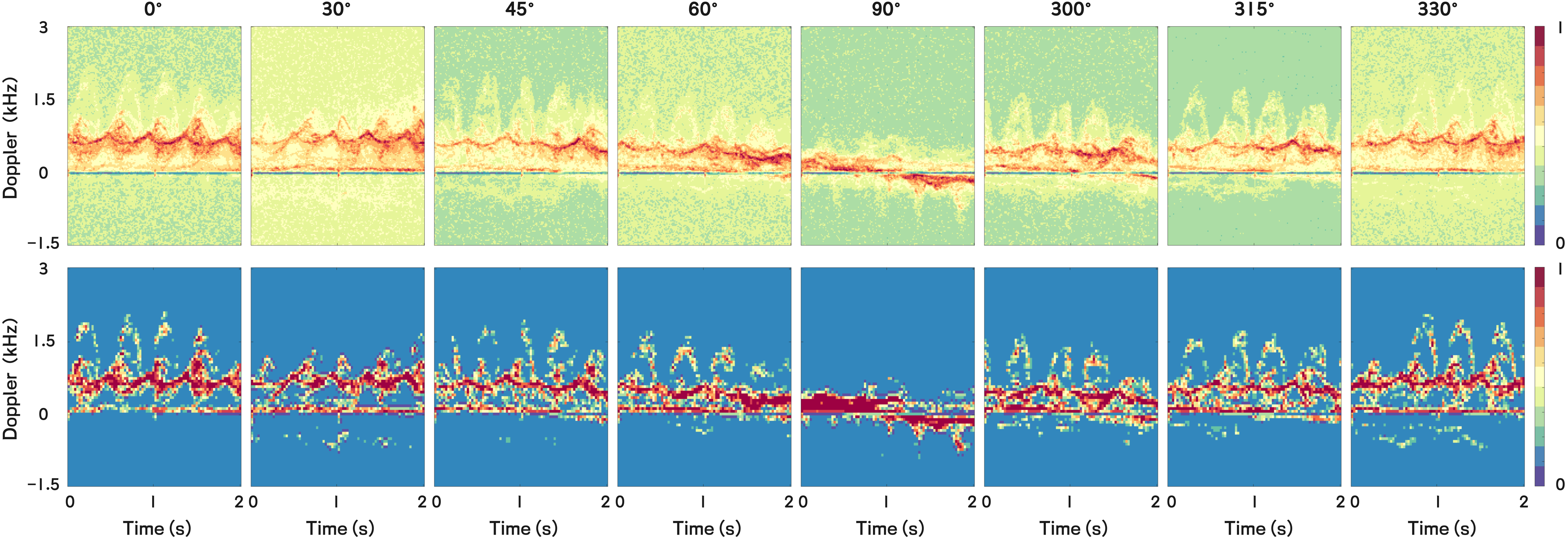}
    \caption{Visualization of DTMs in various motion directions. The first and second row represent the images before and after feature augmentation, respectively.}
    \label{Visualizations}
    \vspace{-0.3cm}
\end{figure*}\par

\section{Low-Cost but Accurate HMDD Method}
As shown in Fig. \ref{HMDD_Network}, the augmented DTM $J$ is input into the proposed HMDD model based on a ViT-CNN hybrid architecture for training and inference \cite{SBCFormer}.\par
The proposed model first applies two layers of $3\times 3$ convolutions with a stride of $2$ to obtain the first feature map, followed by processing through a lightweight feature extraction module. The feature extraction module first uses convolutional embedding. The first path is processed by the InvRes module \cite{MobileNet-V2}, which consists of pointwise convolution (PW-Conv), GeLU activation, and depthwise convolution (DW-Conv). The second path is processed by $7\times 7$ average pooling (Avg-Pool), two InvRes modules, multi-head attention module (MH-Att) \cite{Multi-Head Attention}, transposed convolution (ConvT), and PW-Conv. The results of the two paths are multiplied pointwise and concatenated with the second path to obtain the output of the feature extraction module. Stack the feature extraction module twice more. Finally, global average pooling (GAP), fully connected (Fc), and Softmax activation are used to obtain category labels. Cross-entropy (CE) is used for the loss:\par

\vspace{-0.2cm}
\begin{equation}
\mathcal{L}_{\text{CE}} = -\frac{1}{\mathrm{Bat}} \sum_{b=1}^{\mathrm{Bat}} \sum_{\mathrm{Cla}=1}^{\mathrm{Class}} \mathrm{Tr}_{b,\mathrm{Cla}} \log(p_{b,\mathrm{Cla}}),
\end{equation}
where $\mathrm{Bat}$ is the batch size, $\mathrm{Class}$ is the number of direction categories, $\mathrm{Tr}_{b,\mathrm{Cla}},p_{b,\mathrm{Cla}}$ are the true label and prediction probability of sample $b$ in class $\mathrm{Cla}$, respectively.\par
\begin{figure}
    \centering
    \includegraphics[width=0.48\textwidth]{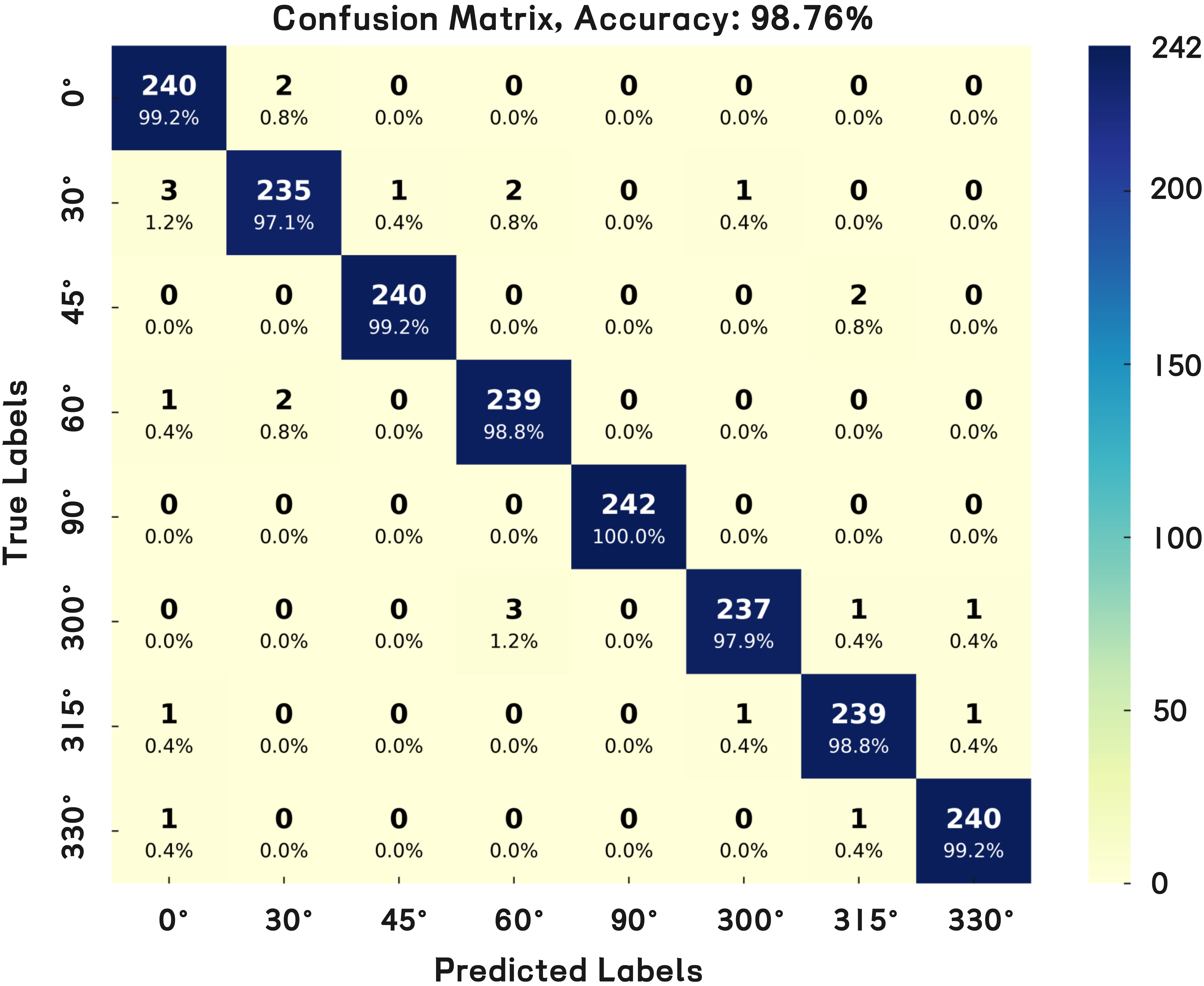}
    \caption{Confusion matrix of HMDD on the validation set.}
    \label{Confusion_Matrix}
    \vspace{-0.4cm}
\end{figure}\par
\begin{figure}
    \centering
    \includegraphics[width=0.48\textwidth]{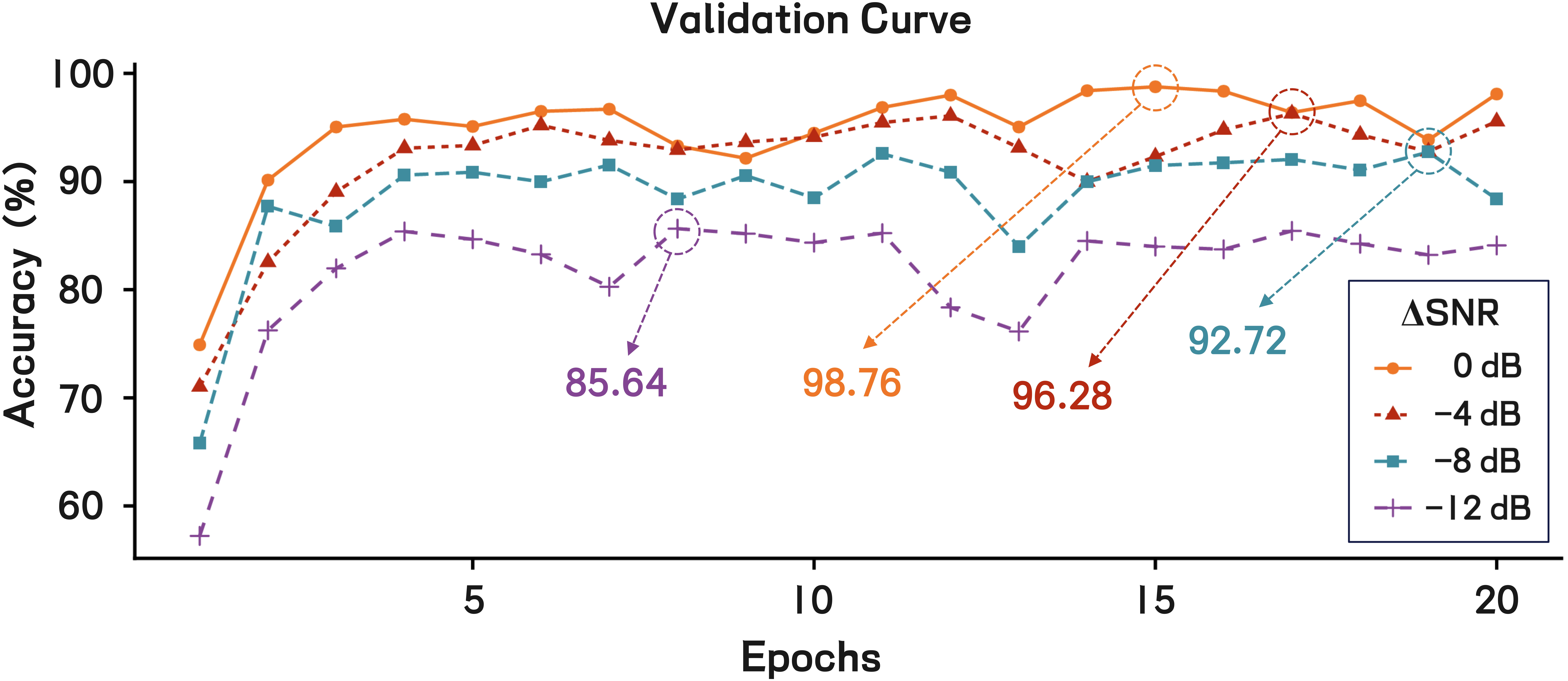}
    \caption{Validation accuracy curves under different SNR conditions.}
    \label{Robustness_Curve}
    \vspace{-0.4cm}
\end{figure}\par

\section{Numerical Experiments}
\subsection{Open-Source Dataset Construction}
The proposed method is experimentally verified using the open-source dataset in paper \cite{Dataset}. The dataset was collected using radar with a center frequency of $77\mathrm{~GHz}$ and a bandwidth of $0.585\mathrm{~GHz}$. The radar has $1$ transmit antenna and $4$ receive antennas. All channels' data is preprocessed and augmented using the methods described in section II. The dataset includes eight different motion direction categories: $0^{\circ}, 30^{\circ}, 45^{\circ}, 60^{\circ}, 90^{\circ}, 300^{\circ}, 315^{\circ}, 330^{\circ}$. The three categories including wearing coats, carrying handbags, and normal walking are combined into a total of $9680$ sets, and randomly divided into training/validation sets at a ratio of $8:2$.\par
The feature maps' dimension of the proposed HMDD network model satisfies $C_1=96,C_2=160,C_3=288$. The total number of training iterations is $20$, with the Adam optimizer, an initial learning rate of $0.00147$, and a batch size of $256$. Feature augmentation and network construction are implemented using MATLAB R2024b and Paddlepaddle 3.0.0 framework in Python 3.10.10. The CPU environment is Intel Core i9-13900H, and the GPU environment is NVIDIA Tesla V100 for training and RTX 4070 Laptop for inference.\par

\subsection{Visualization Verifications}
As shown in Fig. \ref{Visualizations}, visualizations of DTMs in various motion directions are presented. From the first row, it can be seen that micro-Doppler signature appears as a combination of multiple curves on the DTM. The spectral width occupied by micro-Doppler signature varies at different motion directions, but the gait frequency remains relatively consistent. This is a prerequisite for the effectiveness of the proposed feature augmentation method. After feature augmentation, micro-Doppler signature buried in noise is effectively extracted, and the traces of each human node are effectively enhanced. Enhanced features theoretically have better interclass separation. The effectiveness of the proposed feature augmentation method is proved by the above results.\par

\subsection{Accuracy, Robustness, and Cost Verifications}
As shown in Fig. \ref{Confusion_Matrix}, the confusion matrix of the proposed method on the validation set is shown, where the horizontal axis represents the predicted classes and the vertical axis represents the true classes. The proposed method achieved a total validation accuracy of $98.76~\%$, which is high enough for the practical value of HMDD. From the results of the confusion matrix diagonal, it can be seen that the accuracy at $90^{\circ}$ is the best, while the accuracy at $30^{\circ}$ is relatively poor. However, the accuracy of all categories is not less than $97~\%$. Therefore, the proposed method is accurate.\par
As shown in Fig. \ref{Robustness_Curve}, manually add Gaussian noise of varying intensity to DTM, causing the image signal-to-noise ratio (SNR) to decrease by $4\sim 12 \mathrm{~dB}$, and compare the validation accuracy curves of the model training. From the results, the validation accuracy decreases as the intensity of the noise increases. However, within the controllable SNR range, the validation accuracy is not less than $85\%$. The robustness of the proposed feature augmentation and HMDD method is proved.\par
\begin{table}
\begin{center}
\caption{Comparison of Model Parameters and Inference Speed$^{*}$.\label{Cost_Comparison}}
\vspace{-0.1cm}
\resizebox{0.48\textwidth}{!}{
\begin{tabular}{cccc}
\hline\hline
\textbf{Method}  & \textbf{Accuracy ($\%$)}$^{1}$           & \textbf{Param. ($\mathrm{M}$)}    & \textbf{Inference (s)}$^{2}$ \\ \hline
SDP-Net-V1 \cite{SDP-Net} & $96.59$& $2.76$& $0.031$\\
SDP-Net-V2 \cite{SDP-Net} & $98.92$& $7.03$& $0.075$\\
FML-ViT \cite{FML-ViT} & $94.58$ & $2.74$ & $0.029$ \\ 
MobileViTX$^{3}$ \cite{MobileViTX}     & $97.83$ & $5.78$ & $0.061$\\
\textbf{Proposed Method}$^{4}$ & $\mathbf{98.76}$  & $\mathbf{5.33}$ & $\mathbf{0.043}$  \\
\hline\hline
\end{tabular}
}
\end{center}
\footnotesize $^{*}$ There is little existing works dedicated to HMDD. Thus, methods are mainly referenced from the field of activity recognition \cite{Existing Methods}. Simply replace the activity labels with the direction labels to achieve HMDD.\\
\footnotesize $^{1}$ Only validation accuracy is verified in this experiment.\\
\footnotesize $^{2}$ The time required to process a DTM throughout the entire process, including feature extraction and model loading.\\
\footnotesize $^{3}$ The original paper did not provide details on all parameter designs. Therefore, one based on the structure of MobileViT is roughly constructed.\\
\footnotesize $^{4}$ The results include both feature augmentation and HMDD network.\\
\vspace{-0.6cm}
\end{table}\par
As shown in TABLE \ref{Cost_Comparison}, the validation accuracy, number of model parameters, and inference time of the proposed method with existing lightweight methods are compared, including SDP-Net-V1/V2 \cite{SDP-Net}, FML-ViT \cite{FML-ViT}, and MobileNetX \cite{MobileViTX}. The proposed method achieved the highest validation accuracy except for SDP-Net-V2. The model parameter and inference speed of the proposed method are on par with existing lightweight methods. Therefore, the proposed method is an effective attempt for low-cost but accurate Radar HMDD.\par
\begin{table}
\begin{center}
\caption{Ablation Verifications for Method Design$^{*}$. \label{Ablation_Verification}}
\vspace{-0.1cm}
\resizebox{0.48\textwidth}{!}{
\begin{tabular}{ccc}
\hline\hline
\textbf{Method}  & \textbf{Accuracy ($\%$)}$^{1}$  & \textbf{Inference (s)}$^{2}$ \\ \hline
MobileNet-V2 \cite{Existing Methods}        & $92.56$ & $0.032$ \\
MobileViT \cite{Existing Methods}           & $96.95$ & $0.054$ \\
StarNet$^{3}$ \cite{StarNet}                & $95.04$ & $0.049$ \\
Only HMDD Method                            & $93.54$ & $0.013$ \\
\hline
Augmentation + MobileNet-V2                 & $94.58$ & $0.062$ \\
Augmentation + MobileViT                    & $99.33$ & $0.084$ \\
Augmentation + StarNet                      & $98.61$ & $0.079$ \\
\textbf{Augmentation + HMDD Method}$^{4}$   & $\mathbf{98.76}$  & $\mathbf{0.043}$  \\
\hline\hline
\end{tabular}
}
\end{center}
\footnotesize $^{*}$ Ablation verifications are performed for both feature augmentation and lightweight network design.\\
\footnotesize $^{1}$ Only validation accuracy is verified in this experiment.\\
\footnotesize $^{2}$ The time required to process a DTM throughout the entire process, including model loading. Whether feature augmentation is used is indicated in the table.\\
\footnotesize $^{3}$ StarNet-S3 version is needed to ensure good accuracy performance.\\

\vspace{-0.6cm}
\end{table}\par

\subsection{Ablation Verifications}
As shown in TABLE \ref{Ablation_Verification}, ablation verifications are performed for both feature augmentation and lightweight network design. Compared to directly feeding DTM images to a neural network for recognition, the proposed HMDD method has good accuracy, but is not as good as the Mobile-ViT and StarNet with similar model parameters. The proposed HMDD method is the best in terms of inference speed. Feature augmentation method introduces an additional inference time of $0.03~s$ to all recognition networks. After feature augmentation, the accuracy of all recognition networks is effectively improved, with the HMDD method showing the greatest improvement. Overall, the results prove the rationality of the overall design of the proposed method.\par

\subsection{Discussions}
Although the proposed method is attempted for radar-based HMDD and achieved good accuracy with low cost performance. However, there are still some defects in its feature augmentation and network model design, which are worth further exploration and improvement, including:\par
\textbf{(1) Loss of Information by Feature Augmentation:} The time matrix obtained from the iteration of the proposed feature augmentation method is binarized. Thus the magnitude information of the human's nodes is lost. This somewhat reduces the physical interpretability of the proposed method.\par
\textbf{(2) Redundancy Still Exists in Network Design:} The proposed network model still has some redundancy in the input layer, the scale, and the strategy of embedding. In fact the DTM after feature augmentation is not satisfied with the rank. A more efficient way of encoding temporal features remains to be explored.\par

\section{Conclusion}
The micro‑Doppler spectrum width is affected by the human motion direction, thus, valuable prior information for downstream tasks such as gait recognition is provided when human motion direction is determined. However, simultaneous feature augmentation and motion determination have still not been fully achieved by DTM‑based methods. In response, a low‑cost yet accurate radar‑based HMDD method has been explored in this paper. Specifically, radar‑based human‑gait DTMs have first been generated, after which feature augmentation has been achieved with a feature‑linking model. Subsequently, the HMDD has been implemented through a lightweight ViT-CNN hybrid architecture. The effectiveness of the proposed method has been verified with an open‑source dataset.\par



\newpage
\begin{thebibliography}{00}
\bibitem{Main1}M. A. Alanazi, A. K. Alhazmi, O. Alsattam, K. Gnau, M. Brown, S. Thiel, K. Jackson, and V. P. Chodavarapu, “Towards a Low-Cost Solution for Gait Analysis Using Millimeter Wave Sensor and Machine Learning,” \emph{Sensors}, vol. 22, no. 15, pp. 5470, 2022.
\bibitem{Main2}Z. Ni and B. Huang, “Open-Set Human Identification Based on Gait Radar Micro-Doppler Signatures,” \emph{IEEE Sensors J.}, vol. 21, no. 6, pp. 8226-8233, 2021.
\bibitem{Main3}Y. Yang, Q. Mu, B. Li, Y. Ge, Q. Wang and Y. Lang, “Few-Shot Open-Set Gait Recognition Based on Radar Micro-Doppler Signatures,” \emph{IEEE Sens. J.}, vol. 25, no. 13, pp. 25134-25145, 2025.
\bibitem{Main4}Y. Bu, X. Wang, B. Zhang, S. Guo and G. Cui, “Multidomain Fusion Method for Human Head Movement Recognition,” \emph{IEEE Trans. Instrum. Meas.}, vol. 72, pp. 1-8, 2023.
\bibitem{Main5}Z. Chen, G. Li, F. Fioranelli and H. Griffiths, “Personnel Recognition and Gait Classification Based on Multistatic Micro-Doppler Signatures Using Deep Convolutional Neural Networks,” \emph{IEEE Geosci. Remote Sens. Lett.}, vol. 15, no. 5, pp. 669-673, 2018.
\bibitem{Main6}Y. Lang, Q. Wang, Y. Yang, C. Hou, H. Liu and Y. He, “Joint Motion Classification and Person Identification via Multitask Learning for Smart Homes,” \emph{IEEE Intern. Things J.}, vol. 6, no. 6, pp. 9596-9605, 2019.
\bibitem{Main7}W. Li, J. Liu, S. Guo and Y. Jia, “Human Activity Recognition Method Based on Scattering Separation Using Multifrequency Radar Data,” \emph{IEEE Sensors Lett.}, vol. 8, no. 10, pp. 1-4, 2024.
\bibitem{Main8}S. Huan, H. Huang, W. Shang, M. Zhang, X. Yang, and W. Liu, “Improved Vision Transformer Based on Spatiotemporal Micro-Doppler Characteristics for Human Motion Direction Determination,” \emph{IEEE Sensors J.}, early access, 2025.
\bibitem{Jokanovic}B. Jokanović and M. Amin, “Suitability of Data Representation Domains in Expressing Human Motion Radar Signals,” \emph{IEEE Geosci. Remote Sens. Lett.}, vol. 14, no. 12, pp. 2370-2374, 2017.
\bibitem{Ding}C. Ding, H. Hong, Y. Zou, H. Chu, X. Zhu, F. Fioranelli, J. Le Kernec, and C. Li, “Continuous Human Motion Recognition With a Dynamic Range-Doppler Trajectory Method Based on FMCW Radar,” \emph{IEEE Trans. Geosci. Remote Sens.}, vol. 57, no. 9, pp. 6821-6831, 2019.
\bibitem{Song}C. Song, Y. Yang, Y. Lang and C. Hou, “SISO Radar-Based Human Movement Direction Determination Using Micro-Doppler Signatures,” \emph{IEEE Trans. Geosci. Remote Sens.}, vol. 60, pp. 1-14, 2022.
\bibitem{Qiao}X. Qiao, G. Li, T. Shan, and R. Tao, “Human Activity Classification Based on Moving Orientation Determining Using Multistatic Micro-Doppler Radar Signals,” \emph{IEEE Trans. Geosci. Remote Sens.}, vol. 60, pp. 1-15, 2021.
\bibitem{Yang}Y. Yang, J. Li, B. Li, Y. Lang, and W. Zhai, “Few-Shot Omnidirectional Human Motion Recognition Using Monostatic Radar System,” \emph{IEEE Trans. Instrum. Meas.}, vol. 72, pp. 1-14, 2023.
\bibitem{Qi}R. Qi, X. Li, Y. Zhang, and Y. Li, “Multi-Classification Algorithm for Human Motion Recognition Based on IR-UWB Radar,” \emph{IEEE Sens. J.}, vol. 20, no. 21, pp. 12848-12858, 2020.
\bibitem{Jeon}S. Jeon, H. Jeong, J. Lee, and H. Hyun, “Radar Image Extraction Scheme for FMCW Radar-Based Human Motion Indication,” \emph{J. Korean Inst. Electromagn. Eng. Sci.}, vol. 29, no. 6, pp. 411-419, 2018.
\bibitem{Chang}H.-S. Chang, H.-C. Chu, P. Chen, C.-C. Chang, and S.-F. Chang, “Human Motion Analysis Based on Multi-Channel Doppler Radar System,” in \emph{Proc. IEEE MTT-S Int. Microw. Symp.}, pp. 1470-1472, 2019.
\bibitem{Guendel}R. Guendel, M. Unterhorst, E. Gambi, F. Fioranelli, and A. Yarovoy, “Continuous Human Activity Recognition for Arbitrary Directions with Distributed Radars,” in \emph{Proc. IEEE Radar Conf.}, pp. 1-6, 2021.
\bibitem{Waqar}S. Waqar, M. Muaaz, and M. Pätzold, “Direction-Independent Human Activity Recognition Using a Distributed MIMO Radar System and Deep Learning,” \emph{IEEE Sens. J.}, vol. 23, no. 20, pp. 24916-24929, 2023.
\bibitem{HMDD Drawbacks}S. Ahmed, S. Abdullah and S. H. Cho, “Advancements in Radar Point Cloud Processing for Macro Human Movements in Healthcare and Assisted Living Domains: A Review,” \emph{IEEE Sensors J.}, vol. 24, no. 22, pp. 36287-36305, 2024.
\bibitem{Dataset}L. Du, X. Chen, Y. Shi, S. Xue and M. Xie, “MMRGait-1.0: A radar time-frequency spectrogram dataset for gait recognition under multi-view and multi-wearing conditions,” \emph{J. Radars}, vol. 12, no. 4, pp. 892-905, 2023.
\bibitem{FLM}K. Zhan, J. Teng, J. Shi, Q. Li and M. Wang, “Feature-Linking model for image enhancement,” \emph{Neural Comput.}, vol. 28, no. 6, pp. 1072–1100, 2016.
\bibitem{SBCFormer}X. Lu, M. Suganuma, and T. Okatani, “SBCFormer: Lightweight network capable of full-size ImageNet classification at 1 FPS on single board computers,” in \emph{Proc. IEEE/CVF Winter Conf. Appl. Comput. Vis.}, Waikoloa, HI, USA, 2024, pp. 1112–1122.
\bibitem{MobileNet-V2}H. Bechinia, D. Benmerzoug and N. Khlifa, “Approach Based Lightweight Custom Convolutional Neural Network and Fine-Tuned MobileNet-V2 for ECG Arrhythmia Signals Classification,” \emph{IEEE Access}, vol. 12, pp. 40827-40841, 2024.
\bibitem{Multi-Head Attention}Y. Liu, W. Wang, R. Ye, Y. Ren, L. Wang and W. Pang, “Multi-Head Self-Attention-Incorporated YOLOv5s for Satellites Detection,” \emph{IEEE Access}, vol. 12, pp. 167530-167541, 2024.
\bibitem{Existing Methods}I. Ullmann, R. G. Guendel, N. C. Kruse, F. Fioranelli and A. Yarovoy, “A Survey on Radar-Based Continuous Human Activity Recognition," \emph{IEEE J. Microw.}, vol. 3, no. 3, pp. 938-950, 2023.
\bibitem{SDP-Net}Y. Qiu, X. Li, Z. Deng, X. Huang, P. Pan and X. Ma, “A Low-Cost Dual-Path Feature Fusion Network for Omnidirectional Human Motion Recognition Using Monostatic Radar," \emph{IEEE Trans. Aerosp. Electron. Syst.}, vol. 60, no. 6, pp. 7595-7610, 2024.
\bibitem{FML-ViT}M. Ding, G. Dongye, P. Lv and Y. Ding, “FML-Vit: A Lightweight Vision Transformer Algorithm for Human Activity Recognition Using FMCW Radar,” \emph{IEEE Sensors J.}, vol. 24, no. 22, pp. 38518-38526, 2024.
\bibitem{MobileViTX}X. Li, S. Chen, S. Zhang, L. Hou, Y. Zhu and Z. Xiao, “Human Activity Recognition Using IR-UWB Radar: A Lightweight Transformer Approach,” \emph{IEEE Geosci, Remote Sens. Lett.}, vol. 20, pp. 1-5, 2023.
\bibitem{StarNet}X. Ma, X. Dai, Y. Bai, Y. Wang and Y. Fu, “Rewrite the Stars,” in \emph{Proc. IEEE/CVF Conf. Comput. Vis. Pattern Recognit.}, Seattle, WA, USA, 2024, pp. 5694-5703.
\end{thebibliography}
\end{document}